\documentclass[sigconf,nonacm]{aamas}

\usepackage{subfig}
\usepackage{algorithm}
\usepackage{algpseudocodex}
\usepackage{pifont} 
\usepackage{soul,color}
\usepackage{caption}

\algrenewcommand\algorithmicrequire{\textbf{Input:}}
\algrenewcommand\algorithmicensure{\textbf{Goal:}}

\usepackage{balance} 

\newcommand{\mypara}[1]{\vspace*{0.5ex}\noindent{\bf #1}~}

\title[PACCART: Reinforcing Trust in Multiuser Privacy Agreement Systems]{PACCART: Reinforcing Trust in Multiuser Privacy Agreement Systems}

\author{Daan Di Scala}
\affiliation{
  \institution{Utrecht University}
  \city{Utrecht}
  \country{Netherlands}}
\email{daandiscala@hotmail.com}

\author{Pınar Yolum}
\affiliation{
  \institution{Utrecht University}
  \city{Utrecht}
  \country{Netherlands}}
\email{p.yolum@uu.nl}

\begin{abstract}

Collaborative systems, such as Online Social Networks and the Internet of Things, enable users to share privacy sensitive content. Content in these systems is often co-owned by multiple users with different privacy expectations, leading to possible multiuser privacy conflicts. In order to resolve these conflicts, various agreement mechanisms have been designed and agents that could participate in such mechanisms have been proposed. However, research shows that users hesitate to use software tools for managing their privacy. To remedy this, we argue that users should be supported by trustworthy agents that adhere to the following criteria: (i) concealment of privacy preferences, such that only necessary information is shared with others, (ii) equity of treatment, such that different kinds of users are supported equally, (iii) collaboration of users, such that a group of users can support each other in agreement and (iv) explainability of actions, such that users know why certain information about them was shared to reach a decision. Accordingly, this paper proposes PACCART, an open-source agent that satisfies these criteria. Our experiments over simulations and user study indicate that PACCART increases user trust significantly.

\end{abstract}


         \usepackage{graphicx}

\newcommand{\n}[1]{\textbf{#1}}

\newcommand{\angles}[1]{\langle #1 \rangle}
\newcommand{\BibTeX}{\rm B\kern-.05em{\sc i\kern-.025em b}\kern-.08em\TeX}

\theoremstyle{definition}
\newtheorem{definition}{Definition}[section]
\newcounter{obser}
\newtheorem{observation}[obser]{Observation}

\newcommand{\yes}{\checkmark}

\newcommand{\no}{\ding{55}}

\begin{document}

\pagestyle{fancy}
\fancyhead{}

\maketitle

\section{Introduction}

Privacy is the right of individuals to keep personal information to themselves \cite{PP1}. While many systems are built with configurations to enable users to exercise this right, managing privacy is still a difficult problem. On one hand, collaborative systems, such as Online Social Networks and Internet of Things, contain a vast amount of content that pertain to a single individual, making it difficult, if not impossible, for individuals to attend to each piece of content separately~\cite{AS2}. Recent research on privacy agents shows promising results on how agents can help with privacy, such as on detecting privacy violations~\cite{kokciyan2016p}, recommending sharing behavior~\cite{fogues2017sosharp, squicciarini2017tag}, and learning privacy preferences~\cite{kurtan2021assisting,tonge2020image}. An important aspect to consider is \textbf{co-owned} content, such that the content does not belong to a single individual (e.g., medical information), but pertains to multiple people (e.g., a group photo or co-edited document \cite{PT5}). These co-owners of the content can and do have conflicting desires about the usage of the content, leading to what is termed as \textbf{multiuser privacy conflicts (MPCs)} \cite{CS10, PT11}. 

Various decision-making techniques, such as auctions, negotiation, and argumentation have been employed to build systems to resolve MPCs. Simply put, each user that participates in these systems is represented by a privacy agent that knows its user's privacy requirements. The agent participates in the decision-making system on behalf of its user. For auction-based systems, this means bidding on its user's behalf or for argumentation-based systems, this would correspond generating arguments on behalf of its user. Through participation in this system, the agents decide if and how to share co-owned content by resolving conflicts.
Experimental evaluations on these systems yield good performance results. However, it is also known that users have concerns when it comes to using software tools for managing various elements of their privacy~\cite{story2021awareness,jin2022exploring}. Many existing studies of collaborative systems indicate the importance of {\it trust} in making systems usable by individuals~\cite{IOT2,colnago2020informing}. We argue that to realize trust, the privacy agent of a user should satisfy the following properties: 

\mypara{Concealment:} The privacy agent will know the privacy constraints of the user, either through elicitation or learning over time.  When the agent is interacting with others to resolve conflicts, it should reveal as little as possible about these privacy constraints, since the privacy constraints themselves are private information. Therefore, users would know that their privacy is safe with the agent~\cite{IOT2,IOT8}.

\mypara{Equity:} Different users have different privacy stances, in terms of their motivation and knowledge. While some users would fight not to share a piece of content, others will be indifferent. Contrary to some of the existing work in AI that favors users with certain properties~\cite{bias1,bias2}, we do not want any user to be left behind. Ideally, the privacy agent should take the privacy stance of the user into account and be able to help different types of users as equally as possible; thereby creating equity~\cite{PP1,PP10}.

\mypara{Collaboration:} It is possible that a number of agents that participate in the same conflict resolution have similar privacy concerns or complementary information to support a particular privacy decision~\cite{AS11}. Their agents should be able to collaborate in groups. 

\mypara{Explainability:} It is well-studied that often users do not trust privacy tools because of misconceptions~\cite{story2021awareness}. One solution for this is to make the tools explicit to users. But, more importantly, if the agent itself can provide explanations as to why it has taken certain actions, then its user can understand and even configure the agent better for future interactions~\cite{PT2, miller2019explanation}. 

Accordingly, this paper proposes a new Privacy Agent for Content Concealment in Argumentation to Reinforce Trust (PACCART). PACCART can conceal its user's privacy requirements at different levels, while still resolving conflicts. By adapting to different privacy understandings of users, PACCART will provide equitable treatment. At the same time, PACCART will enable agents to work together towards a shared desired outcome. Finally, it will help users understand the actions it is taking. To the best of our knowledge, this is the first privacy agent that brings these desirable properties together. We made PACCART openly available\textsuperscript{\ref{footnote1}}.

The rest of this paper is organized as follows: Section \ref{section2} explains the necessary background theory on argumentation-based agreement systems. Section \ref{section3} formalizes the PACCART model. Section \ref{section4} discusses our realization of the model\footnote{\label{footnote1}\href{https://github.com/PACCART/PACCARTpaper}{https://github.com/PACCART/PACCARTpaper}} and our experimental results. Section \ref{section5} discusses the user study and its results. Finally, Section \ref{section6} systematically compares our approach with related work and gives pointers for future directions.

\section{Background} \label{section2}
We advocate that for an agent to exhibit these four criteria, it is useful to be able to express the relations between privacy preferences in a semantic manner. Thus, as an underlying agreement system, we opt for argumentation as opposed to other decision-making mechanisms such as auctions or negotiation. Below, we review how a privacy agent would use argumentation theory and how by using a dialogical argumentation system it can resolve privacy disputes.

\subsection{Argumentation Theory}\label{argtheory}

Our agent model makes use of argumentation theory for its reasoning. 
We follow the \textbf{structured argumentation} formalism of ASPIC+~\cite{AM12}, but we differentiate between premises and preferences and rename preferences as biases. Thus, an ASPIC+ \textbf{argumentation} or \textbf{dispute} $d=\langle P,R,B,C \rangle$ consists of \textbf{premises} $P = P_o \cup P_n$ (ordinary premises $P_o$ and necessary premises $P_n$), \textbf{rules} $R = R_s \cup R_d$ (strict rules $R_s$ and defeasible rules $R_d$), \textbf{biases} $B = B_p \cup B_r$ (premise biases $B_p$ and rule biases $B_r$) and Contraries $C$.

A dispute is held between two opposing agents, \textbf{proponent} $a_p$ and \textbf{opponent} $a_o$. Agents have access to their \textbf{knowledge base} \textit{KB}, which contains premises, rules and contraries. With this content, agents can form \textbf{arguments}. In order to win the dispute, agents are able to \textbf{attack} each other’s arguments and can \textbf{support} (or \textbf{defend}) their own arguments with subarguments in order to try to win the dispute \cite{AM15}. In some cases an agent is also able to \textbf{forfeit}, giving up on winning the dispute. Arguments can be attacked on their \textbf{weak points}, which is any subargument that is either a consequent of a defeasible rule or any ordinary premise. \textbf{Useful} arguments are arguments that, when added to the dispute, successfully attack any opponent's current arguments. Acceptability conditions of winning or losing are dependent on the chosen \textbf{semantics}. Baroni et al. \cite{AM16} offer an overview of different semantics and their meaning, including \textbf{grounded, preferred, complete} and \textbf{stable} semantics. 

\subsection{Dispute Protocol}\label{background}

In order for an argumentation agent to be able to hold a dispute with other agents about a \textbf{subject}, it follows a communication protocol. The protocol allows agents to \textbf{extend} the dispute, meaning that they take turns adding arguments from their knowledge base to the dispute in order to either defend or attack the dispute subject.

\begin{algorithm}
\caption{Agent Dispute Extension Protocol}\label{alg:protocol}
\begin{algorithmic}[1]
\Require Agents $A = \{a_p,a_o\}$, each with $KB=\angles{P, R, C}$
\Ensure Determine winner of dispute $d$
\State $a \gets a_p$
\While{$d$ is not forfeited}
\If{$a$ can extend $d$} 
    \State $a$ extends $d$
    \If{$a$ is $a_p$}
        \State $a \gets a_o$
    \Else
        \State $a \gets a_p$
        \EndIf
\Else
    \State $a$ forfeits $d$
\EndIf 
\EndWhile
\end{algorithmic}
\end{algorithm}

Argumentation systems like PriArg~\cite{AS1} utilize this kind of extension protocol, as seen in Algorithm \ref{alg:protocol}. According to the extension protocol, if an agent is able to extend the dispute, it does so. An agent extends the dispute by adding any sufficient argument from its knowledge base. Therefore, as soon as an agent is unable to extend the dispute any further, it forfeits the dispute.

The winner of a dispute is determined by evaluating the outcome according to grounded semantics. This way the \textbf{burden of proof} initially lies on the proponent of the dispute, after which agents take turns by extending the dispute until one of them wins. This is done because the agent that initializes the dispute has something to gain by defending the subject.

\section{Model} \label{section3}
 
The PACCART agent consists of a base component, which works similarly to agents in the PriArg system, as it communicates with other agents through a dialogical argumentation framework that follows the same Dispute Extension Protocol, as defined in Section \ref{background}. Following this, four components will be introduced on top of the workings of the base component. 

\subsection{Concealment Component}

In the case of argumentation over privacy issues, the information to be concealed consists of all information that a user’s agent can hold in its knowledge base, including those that pertain to the user's privacy preferences. At any time step, we make a distinction between content that is revealed during a dispute and content that is not (yet) revealed.

\begin{definition}[\textbf{Concealed Rules and Premises}] \label{Concealed Rules}
		Concealment content sets of Agent $A$'s knowledge base $KB$ in Dispute $d$:
		\begin{itemize}
		\item A set of Concealed Rules ${R}_c \subseteq {R}$ with properties: 
		\subitem When Dispute $d$ starts: ${R}_c = {R}$
		\subitem When Agent $A$ uses rule $r$ to extend $d$: $R_c\gets R_c \textbackslash {r}$
   
		\item A set of Concealed Premises ${P}_c \subseteq {P}$ with properties:
		\subitem When Dispute $d$ starts: ${P}_c = {P}$
		\subitem When Agent $A$ uses premise $p$ to extend $d$: $P_c\gets P_c \textbackslash p$ 
		\end{itemize}
	\end{definition}

We make a distinction between content that is concealed and content that is not, by keeping track of different sets throughout the dispute. At the initialization stage of the dispute, agents have not yet shared any content with each other, which means that all content is still concealed. While the dispute develops, each time an agent shares content with another agent to extend the dispute, that content is revealed and therefore removed from the set of concealed content.

We formalize PACCART's concealment component by providing it the ability to adopt a \textbf{privacy behavior}, consisting of three concealing aspects: \textbf{Scope, Division} and \textbf{Dedication}.

\mypara{Scope:} At each point in the dispute, if possible, an agent extends the dispute by adding one or more arguments (Algorithm~\ref{alg:protocol}, Step 4). The amount of useful arguments (as defined in Section \ref{argtheory}) that an agent considers to add at any point of time to the dispute, is called its scope. An agent without any focused scope would add all available useful arguments at once. An agent with a focused scope is able to carefully select a smaller set of arguments, and locally gains control over the amount of the added (and therefore revealed) content. The larger the scope of an agent, the more content is added at each step in the dispute. 

\mypara{Division:} Not all information is equally important. To be able to denote this, 
we split the sets of contents into \textbf{set-families} \cite{brualdi1977introductory} of content. These subgroups can then be ordered to the likings of the agent. This entails splitting the knowledge base into ordered subgroups of different groups of conceal-worthy content. Therefore, based on the original knowledge base $KB = \{P, R, C\}$, we propose an \textbf{ordered subdivided knowledge base ({\textit{OSKB}})}, which includes the following ordered tuples of set-families:
\begin{itemize}
	\item An ordered tuple of premises $O_P = \langle P_1,\dots,P_n\rangle$
	\item An ordered tuple of rules $O_R = \langle R_1,\dots,R_n\rangle$
\end{itemize}

The relation between these ordered set-families $F_X$ and the sets $X$ (with $X = P,R$) all follow the same properties:
\begin{itemize}
\item $\bigcup F_X = X$
\item $\bigcap F_X = \emptyset$
\item $\forall y\in Y, \forall z \in Z  
\big((Y \subseteq F_X \land Z \subseteq F_X) \rightarrow (y = z \leftrightarrow Y = Z)\big)$
\end{itemize}

With the introduced \textit{OSKB}, an agent can order their content based on its concealment preferences. We can therefore treat these two ordered tuples together as one totally ordered knowledge base, subdivided in what we call \textbf{dedication levels}, as follows: 
$L = \langle\{O_{P_1}, O_{R_1}\}, \ldots, \{O_{P_n}, O_{R_n}\}\rangle$. Each level contains one or more premises and rules. The first level $L_1$ contains content at the top of the ordering of each of the \textit{OSKB} tuples, which is the content that the agent is the least concerned about revealing. The last level $L_n$ contains content at the bottom of the ordering, indicating the content that the agent considers most important to conceal and therefore has to fully commit to winning the dispute in order to be willing to reveal these pieces of information. The \textbf{exhaustion} of an agent's division aspect indicates the amount ordered subdivisions an agent makes. The more exhaustive content subdivision, the higher amount of levels an agent splits its \textit{OSKB} up into. 

\begin{figure}
	\centering
	\subfloat[\n{None}\label{l1}]{\includegraphics[width=.225\textwidth]{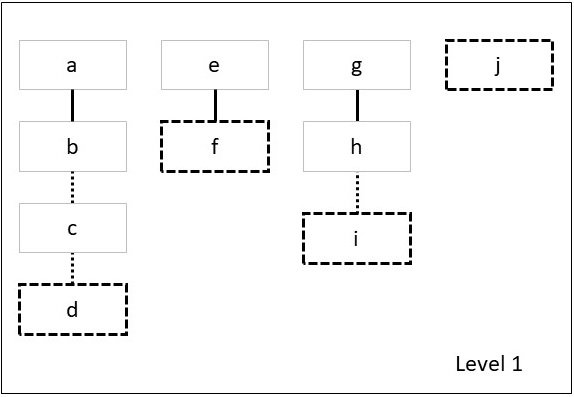}}
	\quad
	\subfloat[\n{Half Arguments} \label{l2}]{\includegraphics[width=.225\textwidth]{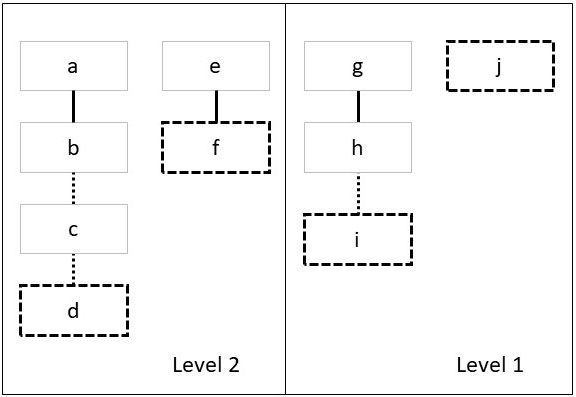}}
	
	\medskip
	
	\subfloat[\n{All Arguments} \label{l3}]{\includegraphics[width=.225\textwidth]{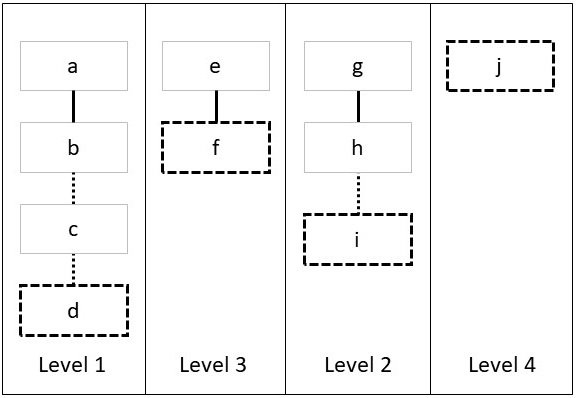}}
	\quad
	\subfloat[\n{All Content} \label{l4}]{\includegraphics[width=.225\textwidth]{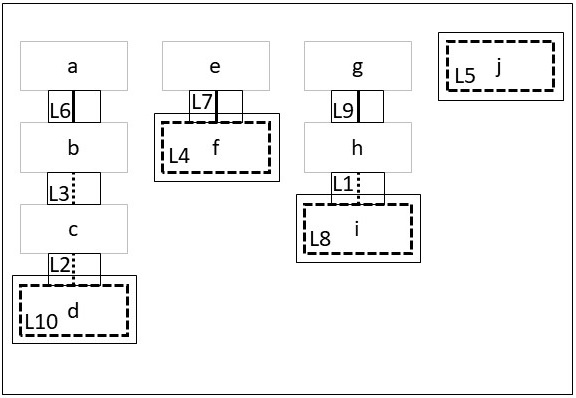}}
	
	\caption{Examples of different approaches of the PACCART agent's division aspect. Four arguments consisting of ten pieces of content are divided up into different levels. Solid and dashed lines are for strict rules $R_s$ and defeasible rules $R_d$, respectively.}
	\label{dividing}
	\Description{This figure shows four examples of different possible divisions of a knowledge base consisting of four arguments of various lengths. In total the arguments consist of ten pieces of content. Pieces of content are marked with letters a through j. In the first example the knowledge base is undivided and therefore it is shown that all content belongs to Level 1. The second example called Half Arguments is shown to be split into two Levels. The third example called All Arguments divides the knowledge base up into four sections, each for one argument. Finally, the example called All Content divides all pieces up into their own level. Note that the levels are assigned randomly, which means that the first piece of content is in Level six and the second in Level three and so on.}
\end{figure}

An example of four different \textit{OSKB} divisions is shown in Figure \ref{dividing}, where an agent makes no subdivision of its \textit{OSKB}  (\ref{l1}), by adding all its arguments to Level 1. Another possibility is it divides its \textit{OSKB} in half, with two levels (\ref{l2}). Furthermore, an agent can choose to divide its \textit{OSKB} in all separate arguments, which yields four levels in this case (\ref{l3}). Note that with this approach, Figure \ref{dividing} shows an example of a level with just one premise ('j'), as only one premise can suffice to form an argument. A final approach consists of an agent dividing its \textit{OSKB} by subdividing all of its content (all rules and premises) over different levels, yielding ten levels in this case (\ref{l4}). 

{\noindent \textbf{Dedication:}} Agents that are able to divide their content into levels, can use this to their advantage. Such agents will initially only provide arguments if they can do so from their first level in their knowledge base. When all arguments in a first level have been depleted, the agents receive the option to either drop to a new level, therefore making a further argument privacy concession, or to forfeit the dispute. This gives agents the ability to weigh their decision to further dedicate to the argumentation. The amount of \textbf{willingness} to drop determines the agent's dedication to continue the dispute. The more willing an agent is to drop dedication levels, the more it will use and therefore reveal the contents of its \textit{OSKB}. This is calculated by whether a certain willingness Threshold $\theta X$ with $X \in [0,100]$ is met at the time of decision whether to commit further to the dispute. This means that an agent with $\theta$75 has a  75\% chance of dropping each level. This entails that the agent example of Figure \ref{l3} has a $0.75^3=42.2\%$ chance to use the content of its final level (as it could drop three times until it reaches its fourth and final level of content), whereas agent example of Figure \ref{l4} has a $0.75^9=7.5\%$ chance to fully commit its \textit{OSKB}.

Any specific combination of all three concealing aspects is called an agent's \textbf{privacy type}. These privacy behaviors are in place for agents to gain additional control over their content concealment during disputes, as well as influence their win rate.

\subsection{Equity Component}

Recall that we want our PACCART agent to be able to help different types of users to deliver on the equity aspect. On user's privacy stances, we follow Dupree et al. \cite{PP2}, who determine a categorization based on stances regarding privacy along two dimensions. We define a user $u$ with \textbf{knowledge} $k \in \{low, medium, high\}$ and \textbf{motivation} $m \in \{low, medium, high\}$. The degree of knowledge indicates the amount of awareness a user has about their privacy and the degree of general knowledge on privacy matters. The degree of motivation indicates the effort a user expends to protect their privacy and the degree of willingness to act on privacy matters. Each system user falls in one of five categories, also known as \textbf{privacy types}:

\begin{itemize}
	\item \textbf{Fundamentalists}: high knowledge, high motivation
	\item\textbf{Lazy Experts}: high knowledge, low motivation
	\item\textbf{Technicians}: medium knowledge, high motivation
	\item\textbf{Amateurs}: medium knowledge, medium motivation
	\item\textbf{Marginally Concerned}: low knowledge, low motivation
\end{itemize}

Dupree et al. determine the rate at which users fall into these categories: 3\% of users are Fundamentalists, Lazy Experts 22\%, Technicians 18\%, Amateurs 34\% and Marginally Concerned 23\%. This is comparable to the categorical distributions of privacy types of earlier conducted researches \cite{ackerman1999privacy,taylor2003most,sheehan2002toward,consolvo2005location,PP8,PP2}.

\textbf{Indifferent} agents are agents that are not personalized and therefore have an unfocused scope and  make no distinction between the importance of content in their \textit{KB}. 
In order for PACCART to be an equitable agent, it should adhere to the following equity properties, which are based on earlier research on equity \cite{PP1,PP10}:
\begin{itemize}
\item[\textbf{EP1:}] 
 The knowledge and motivation of a user is considered and utilized to the fullest extent by their personalized agent.
\item[\textbf{EP2:}] A personalized agent outperforms an indifferent agent. 
\item[\textbf{EP3:}] There are no performance outliers between personalized agents; no personalized agent heavily overperforms or underperforms compared to others.
\end{itemize}

EP1 is important because the strengths of the user should be taken into account by their agent. The privacy stance of a user should not be ignored, as this would be unfair towards users that are heavily engaged in protecting their privacy. In the same line, EP2 is important because the agents that are tailored towards a user should not perform worse than an agnostic, basic agent. Providing personalization should be beneficial for users, not disadvantageous. EP3 is important because in order to reach fair outcomes, it should not be the case that the privacy stance of a user exorbitantly influences the performance of their agent. It would e.g. be unfair towards unknowledgeable users if their agents would underperform by design.

In order to meet these properties, we introduce a mapping between users and agents. This way, both knowledge and motivation are used to determine the personalized agent's privacy type. We determine a fitting mapping between users $u$ to their agents $a$ such that $u \rightarrow a_{\text{scope(shortest)}}$, $u_{\text{knowledge}}\rightarrow a_{\text{division}}$, and $u_{\text{motivation}}\rightarrow -a_{\text{dedication}}$.

First, we assign all personalized agents to have a small scope. This is because a small scope is beneficial for all users, independent of privacy stance. When a user has a high privacy stance, they can let their agent subdivide its content in such a way that each piece of content is thoroughly protected. This would mean that the agent already has a small amount of content to choose from, so for a high privacy user the scope has only a little positive impact. However, in order to also protect users who do not have a lot of knowledge or motivation to bring to the dispute, a small scope is also the best fit in order to protect as much content as possible. 

Secondly, we map a user's knowledge to their agent's division, because of the degree of user knowledge should correspond with the amount of useful subdivisions of their agent's \textit{OSKB} levels. This means that the higher the user's knowledge, the higher the agent's content dividing. Someone with a high knowledge could benefit from an agent with a high capability of dividing its knowledge base content. This would allow users to provide their agent with their preferences in detail. This is in line with EP1. Similarly, mapping a low knowledge to a low \textit{OSKB} division would also be useful. This is because users with low knowledge have little relevant preference divisions to make in their agent’s knowledge base.

Thirdly, we map a user's motivation inversely to their agent’s dedication, because the amount of motivation of a user should correspond to the dedication of its agent to conceal content (in favor of winning disputes). This means that the higher the user's motivation, the lower the agent's dedication. Users that are highly motivated to protect their data would rather have their agent drop as little levels as possible, even if it would require taking (social) losses. Similarly, users that prefer not to act on privacy matters would want their agents to perform well when it comes to winning disputes, but would not mind agents revealing information to do so. This is also in line with EP1.

This mapping results in five personalized agents, one representative for each user type, as noted in Table \ref{nonmps}. This table also includes an indifferent agent.

\begin{table}

	\caption{All three concealing aspects of indifferent PACCART agent and personalized PACCART agents that are matched with representative agents for different user privacy types.}

\begin{tabular}{cccc}
\toprule
                              \textbf{Privacy Type}& \textbf{Scope} & \textbf{Division} & \textbf{Dedication} \\ \midrule
\textbf{\textit{Indifferent}} & All          & None         & $\theta100$         \\ 
\textbf{Fundamentalist}            & Shortest     & AllContent   & $\theta25$          \\ 
\textbf{Technician }          & Shortest     & AllArgs      & $\theta25$          \\ 
\textbf{Amateur}              & Shortest     & AllArgs      & $\theta50 $         \\ 
\textbf{Lazy Expert}          & Shortest     & AllContent   & $\theta75   $       \\ 
\textbf{M.Concerned}          & Shortest     & HalfArgs     & $\theta75 $         \\ 
\bottomrule
\end{tabular}

\label{nonmps}
\end{table}

\subsection{Additional Usability Components}

In addition to the Concealment and Equity components two usability measures are taken. A Collaboration component is introduced to support both sides of the dispute to be represented by multiple agents. This is achieved by introducing the notion of teams such that the set of agents $A$ in the protocol now consists of $A = \{T_p, T_o\}$ to support both a proponent team $T_p=\{a_{p1},\ldots,a_{pn}\}$ and opponent team $T_o =\{a_{o1},\ldots,a_{on}\}$. In order to extend a dispute each team of PACCART agents continuously selects one of its agents to extend. A team forfeits when none of its agents can extend the dispute any further. This component allows for multiple PACCART agents to cooperate on a common goal of defending/attacking a privacy related subject. This means that agents can add content from their own \textit{OSKB} to the dispute when other agents in their team fail to do so.

Furthermore, an Explainability component is introduced to give users insights to the working of their agent. The semantic nature of PACCART allows us to produce both textual and visual output. PACCART can provide textual output by considering outcomes and providing feedback to the user. Based on this, it is able to give different kinds of feedback, with a range of detail. It can notify users on a summary (e.g., \textit{"I have won 56\% of today's disputes and managed to conceal 73\% of your content"}) or it can give detailed advice on possible actions to be taken to improve its performance (e.g., providing its user with a list of weak arguments to remove or strengthen). Furthermore, PACCART can provide visual output by showing its user images of the Structured Argumentation Framework \cite{AM12} of final disputes. This gives users a visual overview of (counter)arguments and possible weak points in their content. This component allows users of PACCART to better understand its inner workings and performance.

\section{Experimental Results}  \label{section4}

The PACCART agent and the experimental setup are implemented as a C\# program. For the sake of reproducibility, we make this program and experiments open source\textsuperscript{\ref{footnote1}}.

\subsection{Dataset Generation}

We implement a system that generates datasets of disputes according to four parameters. The \textbf{disputeAmount} parameter indicates the amount of generated unique disputes that adhere to the other parameters. A higher input value indicates a larger set of disputes, therefore less prone to outliers. The \textbf{disputeSize} parameter controls the amount of arguments that the dispute can contain. A higher input value indicates larger disputes with more content. The	\textbf{maxArgumentSize} parameter dictates the maximum amount of subarguments that each argument can consist of. A higher input value indicates larger arguments with more content and therefore more attackable weak points. Finally, \textbf{maxBranches} is used to control the maximum amount of attacks that each weak point can have, indicating a branching choice in the dispute. A higher input value indicates more options for both agents.

By tuning these parameters, we are able to generate dispute datasets of various shapes and sizes, which makes for exhaustive possibilities for testing functionalities of PACCART. 
After preliminary analysis of variables, we generate a dispute dataset based on the default parameter settings (disputeAmount = 200, disputeSize = 20, maxArgumentSize = 10, maxBranches = 2).

\subsection{Experiment 1: Effect of Privacy Behaviors}

\subsubsection{Setting}

The goal of the first experiment is to test the performance of PACCART agents. Agent performance is evaluated on two metrics, average concealment $C_{avg}$ and average win rate $W_{avg}$. We hypothesize the following:
\begin{itemize}
\item[\textbf{H1:}] A smaller scope leads to both increased concealment and increased win rate.
\item[\textbf{H2:}] More exhaustive division leads to increased concealment and decreased win rate.
\item[\textbf{H3:}] A higher dedication leads to decreased concealment and increased win rate.
\end{itemize}

We determine four or five conditions for each of the three privacy behavior aspects, to test the range of PACCART's concealing behaviors. For the scope, we include selecting the \n{Shortest} or \n{Longest} arguments, as well as a \n{Random} argument or \n{All} possible arguments. For the division, we follow the examples of Figure \ref{dividing} and include conditions where \n{None} of the content is split, where the \textit{OSKB} is split into two groups of arguments (\n{HalfArgs}), split into all separate levels of arguments (\n{AllArgs}) or a subdivision where each level contains a single piece of content (\n{AllContent}). The dedication conditions consist of an increasing threshold $\theta$, with $\theta \in \{0, 25, 50, 75, 100\}$ that should be met in order to drop to a new level. These conditions yield 80 possible privacy types. Each of these 80 predetermined agents are set up against all other agents, and simulations are run on the 200 disputes of our dataset. This means that the experiment is run on 16,000 disputes for 80 agent set-ups, totaling in 1,280,000 simulated disputes. For each of the disputes, both agents are evaluated as a proponent, as well as opponent of the dispute, to ensure equal chances of winning. 

\subsubsection{Results}

\begin{figure}[H]
	\centering
	\includegraphics[width=\linewidth]{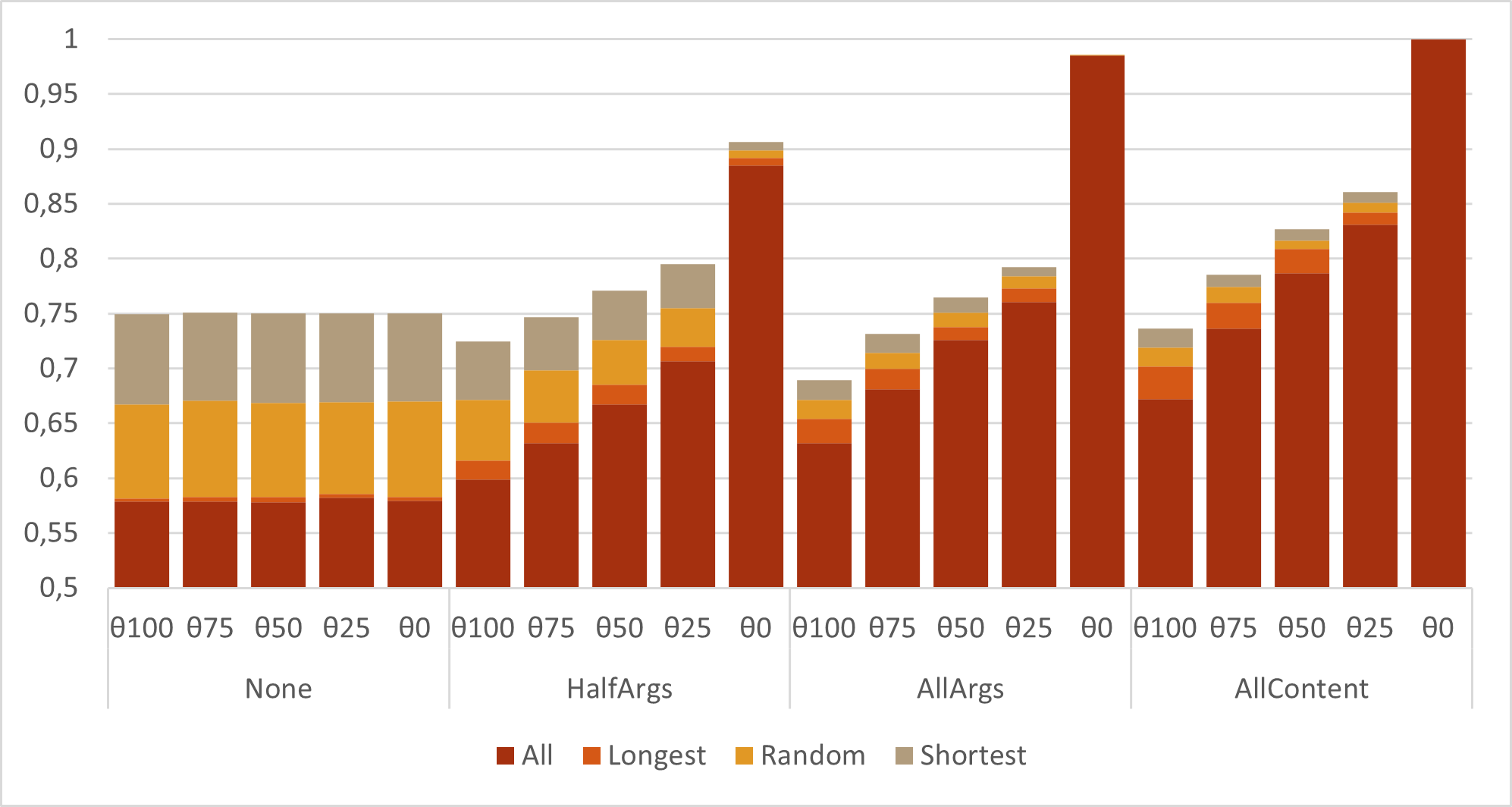}
	\caption{Average Concealment $C_{avg}$ results for all PACCART privacy types.}
	\label{Graph:Concealment3a}
	\Description{Grouped bar graph showing the concealment results. The ranges are typically between 0.55 and 0.8, with the exception for the theta 0 results, which ranges between 0.55 and 1. The theta results for None division are all the same. For the other division types, an upward trend is shown.}
\end{figure}

\begin{figure}[H]
	\centering
	\includegraphics[width=\linewidth]{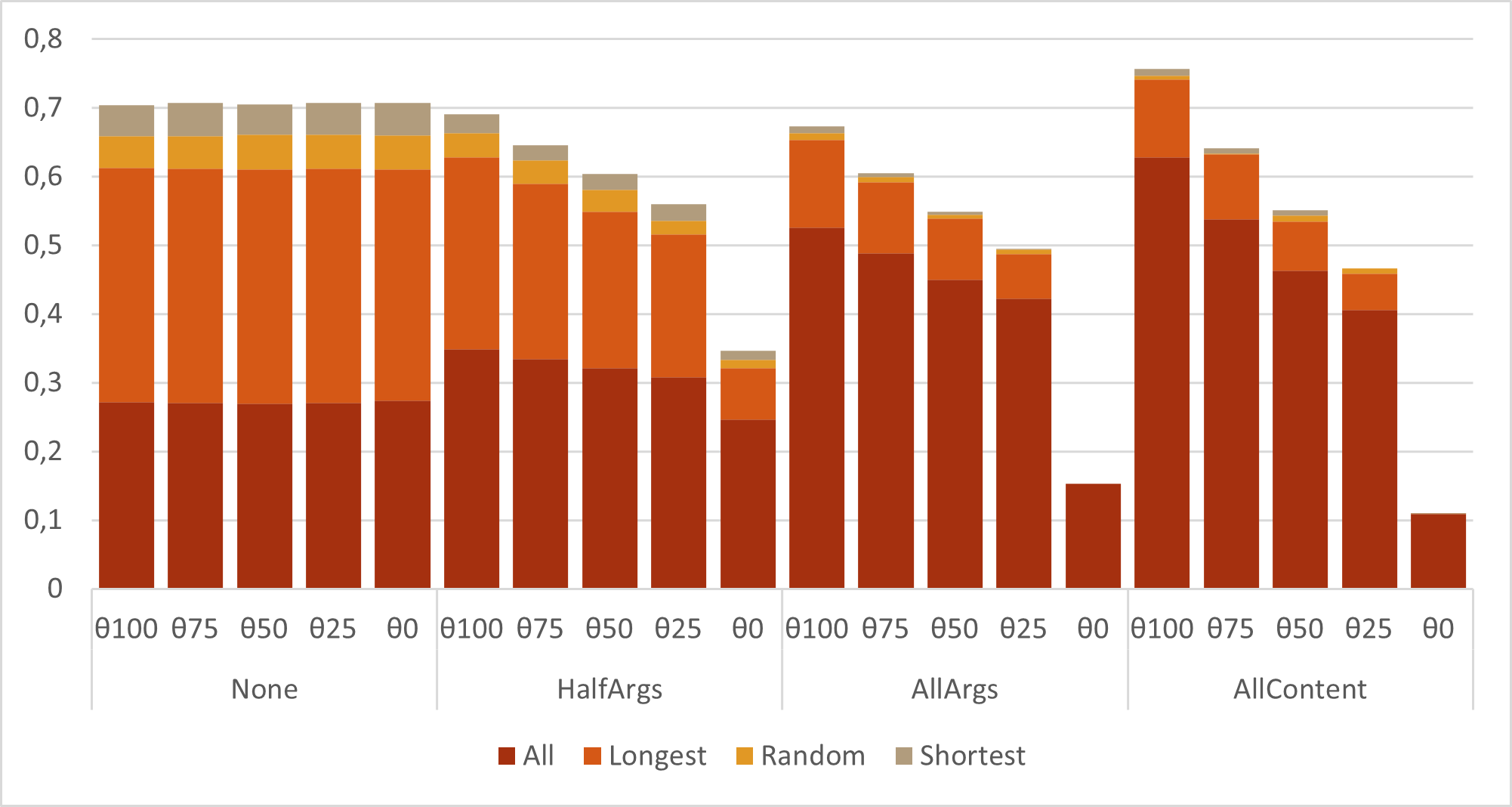}
	\caption{Average Win Rate $W_{avg}$ results for all PACCART privacy types.}
	\label{Graph:Concealment3b}
	\Description{Grouped bar graph showing the win rate results. The ranges are typically between 0.3 and 0.7, with the exception for the theta 0 results, which ranges between 0.1 and 0.7. The theta results for None division are all the same. For the other division types, a downward trend is shown.}
\end{figure}

Figures \ref{Graph:Concealment3a} and  \ref{Graph:Concealment3b} depict the performance of the 80 different agent privacy behavior types, across all three concealing aspects.

{\noindent \textbf{Scope:}} We observe from Figures \ref{Graph:Concealment3a} and  \ref{Graph:Concealment3b}  that the scope of an agent has a significant effect on its performance. Both the average win rate $W_{avg}$ and average concealment $C_{avg}$ increase with a smaller scope. We conclude that a smaller scope has a strictly positive impact on both metrics. This confirms hypothesis H1.

\mypara{Dividing:} All of the \textbf{None} dividing aspect results are equal, independent of dropping willingness. This means that not dividing the \textit{OSKB} negates the effect of the agent's dedication. This is an expected outcome, which happens because there is no division made of the knowledge base so there are no levels for the agent to drop between, even if it would be willing. Outside of this behavior, an upward trend is noticeable in all cases for average concealment, as well as a downward trend in all cases for the win rate, with more exhaustive dividing. This confirms hypothesis H2.

\mypara{Dedication:} When looking at the dedication aspect, we observe an upward trend in all cases for average concealment $C_{avg}$, as well as a downward trend in all cases for win rate $W_{avg}$, with less willing dedication. This is a similar trend as with the dividing aspect of the privacy behavior. This confirms hypothesis H3. Furthermore, Figure \ref{Graph:Concealment3b} shows a significant drop in win rate from $\theta$25 to $\theta$0, while the improvement in concealment is disproportional. This shows that it is beneficial for an agent to be at least somewhat willing to commit to the dispute.

Based on these results, we conclude the following observation:  

	\begin{observation}\label{Observation 3}
		\textit{PACCART's concealment component allows users to  keep information private, while also giving them the choice of a trade-off between winning disputes and further protection of information.}
	\end{observation}

\subsection{Experiment 2: Effect of User-Agent Mapping in Realistic Setting}
\subsubsection{Setting}

The goal of the second experiment is to evaluate the mappings between agents and users by simulating disputes for each personalized agent in a realistic setting. The results of this mapping will determine whether EP2 and EP3 are met, which means that PACCART is an equitable agent. Therefore, based on this mapping, we further hypothesize:

\begin{itemize}
\item[\textbf{H4:}] Equity property EP2 is met under a mapping where personalized agents are assigned the smallest possible scope.
\item[\textbf{H5:}]  Equity property EP3 is met under a mapping where personalized agents are assigned a fitting trade-off between division and dedication.
\end{itemize}

We create a set of opponents according to data of distribution of real life user population as given by Dupree et al. This opponent set therefore contains three Fundamentalist agents, 22 Lazy Expert agents, 18 Technician agents, 34 Amateur agents and 23 Marginally Concerned agents. We call this set of 100 agents the \textbf{Model Population Set MPS}. The MPS is in place because in a practical scenario it is less likely that an MPC occurs between Fundamentalists' agents, as between Marginally Concerned users' agents\footnote{An additional experiment is performed to evaluate the MPS, placing all agents in a non-distributed setting, which yielded similar results.}.

This means that six agents (one indifferent agent and all five personalized agents) compete 100 times against each of the personalized agents, and simulations are run on 200 disputes on the dispute dataset. Overall, the experiment is run on 20.000 disputes for six agent set-ups. Furthermore, agents are again tested twice for all disputes, both as proponent and opponent of the subject, to ensure equal chances of winning. 

\subsubsection{Results}
\begin{figure}[H]
	\includegraphics[width=\linewidth]{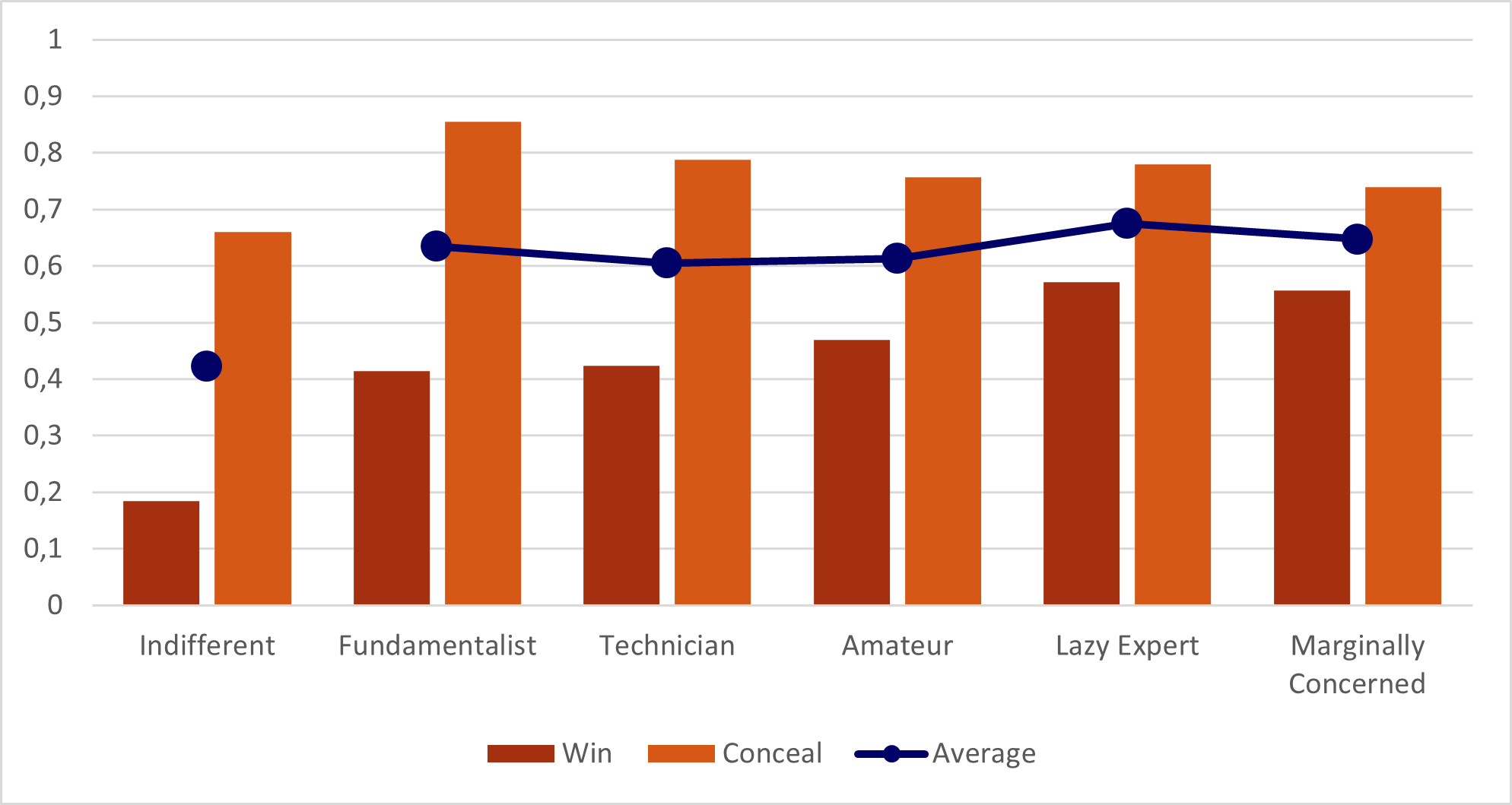}
	\caption{Average win rate $W_{avg}$ and Average Concealment $C_{avg}$ for indifferent agent and personalized agents in MPS. Averages between $W_{avg}$ and $C_{avg}$ are indicated with a line.}
	\label{graphs equity2} \Description{Grouped bar graph showing the results for all six PACCART agents. The win rate ranges for personalized agents between 0.4 and 0.6. The concealment ranges for personalized agents between 0.9 and 0.7. An average line is shown to be between 0.6 and 0.7 for all personalized agents. Results of an indifferent agent is shown as a side by side comparison, with significantly lower performance.}
\end{figure}

The results of the second experiment can be seen in Figure \ref{graphs equity2}. Again, performance is measured by concealment $C_{avg}$ and win rate $W_{avg}$. As shown in Figure~\ref{graphs equity2}, the indifferent agent performs much worse than the personalized agents on both metrics (only 0.185 for win rate and 0.660 for concealment). This confirms hypothesis H4.

Furthermore, the averages of all personalized agents are in the range between 0.6 and 0.7. This means that although some personalized agents are better at winning or concealing, the overall performance leads to an equitable situation where no users are victimized by the agent's workings. This confirms hypothesis H5.

It is worth noting that an interesting trend occurs between personalized agents, where the Fundamentalist representative’s agent (with the highest privacy stance) wins the least and conceals the most, while the Marginally Concerned representative’s agent wins the most and conceals the least. This trade-off shows how the different privacy stances influence the results. 

Based on these results, we conclude the following observation: 

	\begin{observation}\label{Observation 5}
		\textit{PACCART's equity component allows for a well-matched personalization for users of various privacy stances. While personalized PACCART agents overall perform relatively well, a consistent trade-off between win rate and concealment shows that no user is disadvantaged.}
	\end{observation}

\section{User Study}  \label{section5}
We further conduct a user study to understand what components of PACCART lead to user trust. 
\subsection{Setting}
We design a survey in two parts. The first part of the survey has questions on the privacy stance of participants, in order to assess their privacy type. We deliberately use existing questions from the literature to ensure compatibility: three questions used by Westin et al. (e.g., \textit{"How much do you agree with the statement `Most businesses handle the personal information they collect about consumers in a proper and confidential way.'?"})~\cite{PP3} to determine the knowledge of participants on privacy and 10 questions on statements about privacy from the study of Dupree et al., to determine the motivation of participants on privacy (e.g., \textit{"How strongly do you identify yourself with the statement `I would rather choose being social over privacy.'?"}). As validation and to mitigate response bias, we also ask participants directly to self-assess their own knowledge and motivation (e.g., \textit{"How much do you know about digital privacy issues?"}). These questions are all answered on a Likert scale. The full questionnaire is also made openly available\textsuperscript{\ref{footnote1}}.

The second part of the survey has questions on the various components of PACCART as a personal assistant. This part starts with an example scenario. Then a set of questions follows in which participants are asked to rate their perceived trust of such personal assistants on a Likert scale (1 = Strongly Distrust, 5 = Strongly Trust). The first question is on the participants’ initial thoughts of trust on the PACCART base component (an explanation followed by \textit{"How much would you trust to use such a privacy assistant?"}). Then, each separate PACCART component is explained separately and addressed as a question. Afterwards, the participants are asked to rate the agent with all components combined (the base component with all four additional components). Finally, the participants are asked to reconsider their thoughts on the base component. This gives the participants a chance to reflect on their initial thoughts.

The survey is distributed through Qualtrics, an online, secure cloud-based, survey tool. Data is automatically and anonymously recorded through Qualtrics, in accordance with GDPR requirements. The survey is preceded by filling out a consent form. To ensure correctness and clarity, we first perform a small pilot study. Afterwards, the survey is distributed online for a user study. The first part of the survey is used to gather participants for a final interview study, in which we collect opinions on the agent by participants of various privacy stances.

\subsection{Results}
\subsubsection{Pilot Study}
 Data and feedback was collected from three participants in the pilot study. Each of these participants were categorized as a different privacy type: one Lazy Expert, one Technician and one Amateur. They had no trouble with filling the survey and found the Qualtrics interface to be non-distracting. However, some feedback was given on the wording of questions and options to answers. We updated the wording of the second part accordingly to avoid ambiguity and then began the actual study. 

\subsubsection{User Study}
Data was collected from $117$ voluntary participants in the user study. Based on validation questions and completion requirements, $12$ survey responses are filtered out. Out of the remaining $105$ participants, eight participants self-assessed as Fundamentalists, $20$ participants as Lazy Experts, $22$ as Technicians, $31$ as Amateurs and $24$ as Marginally Concerned users. This is in line with the distributions by Dupree et al. ~\cite{PP2}.

 We report the mean (M) and standard deviation (SD) of the results, as well as significance through t-tests (\textit{P}). The results indicate that the initial consideration of the PACCART base component is fairly neutral (M = 2.857, SD = 1.023), slightly leaning towards distrust. The trust ratings given by participants are higher than the initial consideration for both Concealment (M = 2.943, SD = 0.979) as well as Equity (M = 3.171, SD = 1.069). There is a significantly ($P<.001$) positive increase of trust of the combined agent (M = 3.467, SD = 0.974) compared to the initial consideration of the base component. Even more so, when asked to reevaluate the trustworthiness of the agent, the average trust rating significantly ($P<.001$) drops (M = 2.362, SD = 0.982) compared to the combined agent. These results strongly indicate that overall, the principle of PACCART and its components increases the indicated trust of users.

 \begin{figure}[H]	\centering
	\includegraphics[width=\linewidth]{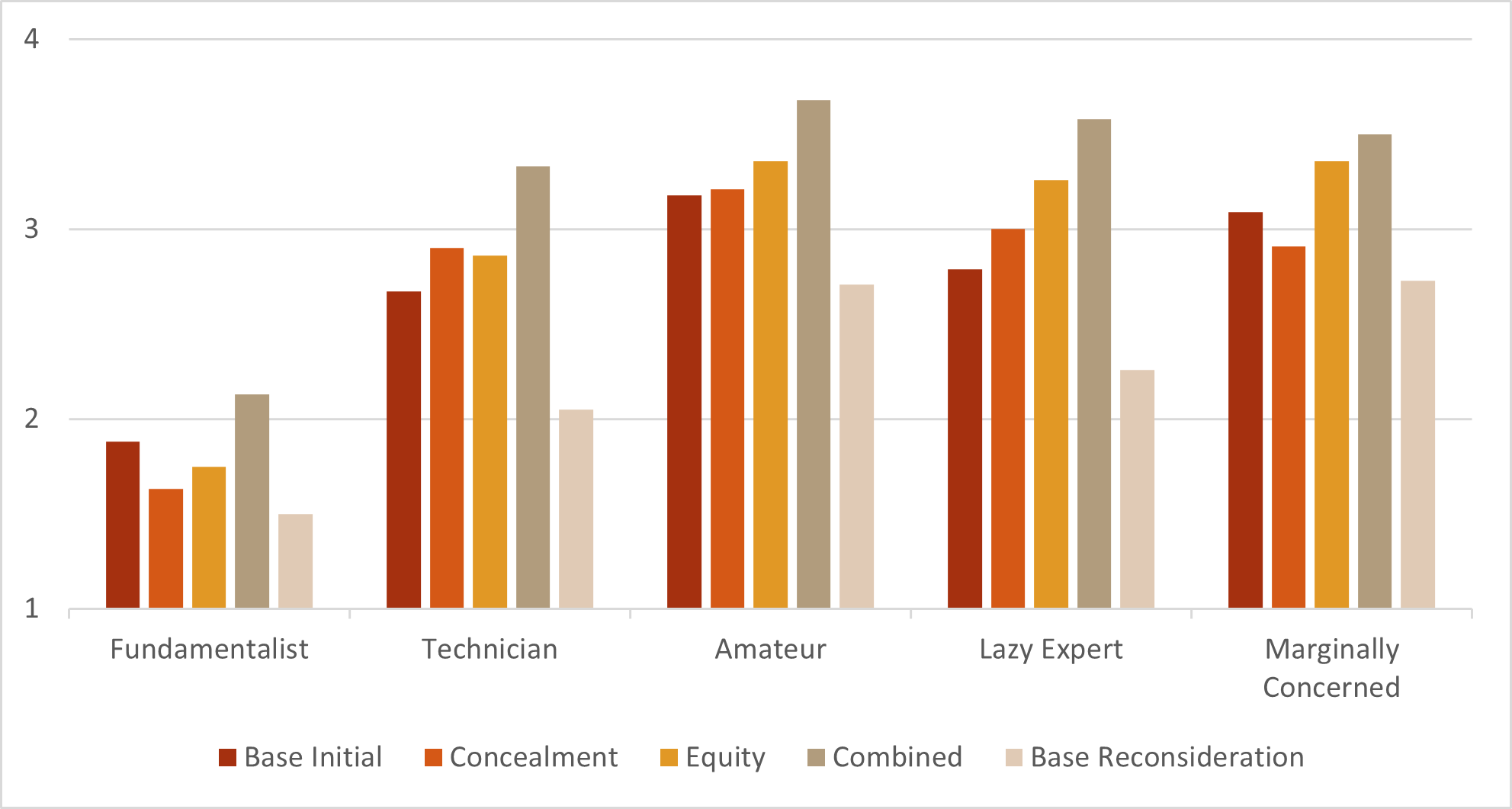}
	\caption{Average survey study ratings of participant trust on a Likert scale (1 = Strongly Distrust, 2 = Distrust, 3 = Neutral, 4 = Trust, 5 = Strongly Trust), divided by privacy type.}
	\label{Graph:survey}
	\Description{Grouped bar graph showing the results for all five participant privacy types. Typically, the Fundamentalist scores are much lower than the other scores, scores ranging between 1 and 2.5 whereas for example marginally concerned results are typically shown to be around 3.}
\end{figure}

The results are split on each of the privacy type categories, as shown in Figure \ref{Graph:survey}. The graph shows the average trust ratings by users with different privacy types. From this graph, we can observe the effect of different stances on privacy on agent trust scores. These results show that the lower the privacy stance, the higher the overall trust in the agent. Fundamentalist participants' highest mean trust rating is a 2.130, whereas Marginally Concerned participants’ lowest mean trust rating is a 2.727. This is in line with our expectations about the privacy types and therefore an indication that the privacy stance assessment part of the survey works as intended.

A further noteworthy observation is that for all of the individual privacy types the reconsideration is rated lower than the initial thoughts on the base component. This indicates that after having read an explanation on what possible components could improve upon the base, participants independently of their privacy type believe the base component to be less trustworthy. When comparing the base component with the total combined agent, trust significantly increases for all user types ($P < .001$) except for Fundamentalists. While the results do indicate an increase of trust for Fundamentalists, the results are not significant ($P= .18$), which is expected because of the naturally low occurrence of users with this high privacy stance. These results strongly indicate that overall, the principle of PACCART and its components increases the indicated trust of users of all privacy stances.
    
\subsubsection{Interview Study}
Data was collected from eight participants in the interview study. To ensure a well-balanced feedback from the interview participants, a distribution of participants with varying stances is selected. The participants consisted of one Fundamentalist (F1), two Lazy Experts (L1, L2), one Technician (T1), two Amateurs (A1, A2) and two Marginally Concerned users (M1, M2). 

The results from the interview study follow the same trend as the results from the user study. The Marginally Concerned participant M2 remarked that they would trust to use PACCART because if they would not use the agent at all, they would have less control about protecting their privacy. T1 and L1 indicated that the agent's concealing behavior deeply increased their trust, as this would increase their control over their privacy. F1 and L1 responded more warily, and indicated that trusting such an agent highly depended on the trustworthiness of its producers, and that their opinion of the agent would depend on the reception by other people within their belief system. A2 indicated that an equitable agent would increase their trust. However, they would find it difficult to understand whether the personalization would be correct for them. This points us to possible future work in which agents learn from their user.

Furthermore, L1 indicated that the possibility of collaboration increased their trust a lot because they could divide their information across multiple agents within different devices. L2 and A2 indicated that feedback from the agent increased trust, as well as would heighten their awareness about their privacy. Finally, many participants (L2, T1, A1) indicated that they would trust to use the fully combined agent, especially if through future additions they would be able to improve its personalized concealing behavior by using the provided feedback to fine-tune their system.  

\balance 

\section{Discussion}  \label{section6}

We first evaluate our approach with respect to three leading approaches that provide privacy agents for MPCs and then discuss future directions in which our work can improve.
\subsection{Comparison}

\begin{table}\caption{Comparison of privacy criteria between approaches.}\label{comparison}
\begin{tabular}{ccccc}
\toprule
 Approach & CON & EQU & COL & EXP \\ 
 \midrule
PACCART & \yes & \yes & \yes & \yes \\ 
PANOLA & \yes & \yes & \no & \no \\ 
ELVIRA & \no & \no & \yes & \yes \\ 
Filipczuk et al.  & \no & \yes & \yes & \no \\ \bottomrule
\end{tabular}
\end{table}

PANOLA \cite{PP1} is an agent that participates in an auction system for privacy. It incorporates privacy types for personalization. PANOLA can learn to bid correctly in order to optimize their privacy preservation. ELVIRA \cite{mosca2021elvira} is a practical reasoning agent, which is designed for collaborative resolutions of MPCs. ELVIRA is both value- and utility-driven, and is able to produce an explanation of its process. The system by Filipczuk et al. \cite{filipczuk2022automated} is a multi-issue negotiation framework, designed to learn from its users' preferences. This system allows users to focus on the privacy issues that they find important, by including the users in the outcome decision loop. 

We perform a comparison based on the earlier defined desirable properties for trustworthy assistive MPC approaches: concealment (CON), equity (EQU), collaboration (COL) and explainability (EXP). This comparison is summarized in Table \ref{comparison}.

\mypara{Concealment.} PANOLA conceals the privacy constraints as it operates on an auction system and thus provides bids. Thus, there is no dialogue system that would make it possible to reveal privacy constraints. The approach by Filipczuk et al. uses negotiation to achieve a desired MPC solution and does not provide any particular emphasis on concealment. The formulation of offers can easily reveal privacy constraints of the user. ELVIRA works under the epistemic assumption that agents share the same knowledge, where the uploader agent resolves the conflict for all, thus concealment is not possible. While PACCART shares content through argumentation, with its concealing behavior PACCART has control over which content to share with opposing agents. 

\mypara{Equity.} ELVIRA is role-agnostic and thus eliminates the possibility of agents' acquiring advantages through roles. However, it does not provide any explicit treatment to show that the proposed agent helps different types of users well. Filipczuk et al. provide personalization through use of privacy types, which could lead to equity. Both PACCART and PANOLA specifically tailor to different user privacy types and demonstrate how equity is achieved.

\mypara{Collaboration.} PANOLA runs on an auction, on which agents are meant to participate independently. Thus, it does not provide collaboration in groups of agents.
ELVIRA resolves conflicts in a centralized manner. Since all information is shared, the groups are inherent in the system. Filipczuk et al. do not treat collaboration separately, but their use of negotiation could cater for collaboration among groups of agents. PACCART enables agents to form teams in order to compete against others in the system.

\mypara{Explainability.} Where auctions and negotiations are held with bids and offers that can be based on weights and scores, argumentation and reasoning include a way of justification and meaning in their arguments. Therefore, the semantic nature of PACCART and ELVIRA allows for these agents to report back to their users and explain their workings. Neither PANOLA nor Filipczuk et al.'s approach include feedback to the user.

\subsection{Conclusion \& Future Work}
We introduced PACCART, which helps users preserve privacy by enabling automated privacy argumentation.
PACCART aims to induce trust by increasing 
content concealment, providing equitable personalizations, enabling multiagent team-based collaboration and explaining its actions through feedback. The agent is designed to be general and is made publicly available as an open-source program together with the dispute dataset generation system, so that they can be used for research as well as in practical applications, such as team collaboration tools (e.g., MS Teams) where co-owned data is shared abundantly and privacy disputes need to be resolved.

Future research could further build upon closing the feedback loop between users and the agent to further increase trust. When users get prompted that their agent lost a dispute because of the lack of arguments, the user could respond by taking action to help and improve the agent fit to its user. Furthermore, introducing mutual feedback opens new possibilities for machine learning approaches. Now, there exists a mapping between users and their personalized agents, which could be changed into the agent learning the preferences of the user instead. Weights could be given to the importance of dedication to win certain disputes, or concealing specific levels of content. The inclusion of reinforcement learning could be a great additional step towards robust and well-adjusted argumentation based privacy assistants.

\bibliographystyle{ACM-Reference-Format} 
\bibliography{bibliography}

\end{document}